\documentclass[10pt,conference]{IEEEtran}
\IEEEoverridecommandlockouts

\usepackage[utf8]{inputenc}
\usepackage[T1]{fontenc}
\usepackage{microtype}

\usepackage[misc]{ifsym}
\usepackage{mdwlist}
\usepackage{graphicx}
\usepackage{textcomp}
\usepackage{xcolor}
\usepackage{subfigure}
\usepackage{listings}
\usepackage{xcolor,pifont}
\usepackage{url}
\usepackage{balance}
\usepackage{xspace}
\usepackage{bbding}
\usepackage{xcolor}
\usepackage{pifont}
\usepackage{booktabs}
\usepackage{array}
\usepackage{algorithm}
\usepackage{paralist}
\usepackage{multirow,makecell}
\usepackage{enumerate}
\usepackage{fancynum}
\usepackage{algorithm}
\usepackage{algorithmic}

\newcommand{\tabincell}[2]{\begin{tabular}{@{}#1@{}}#2\end{tabular}}

\newcommand{\latinlocution}[1]{\textit{#1}}
\newcommand{\eg}{\latinlocution{e.g.,}\xspace}
\newcommand{\ie}{\latinlocution{i.e.,}\xspace}

\newcommand{\eat}[1]{}

\newcommand{\haoyu}[1]{{\color{red} HW: #1}}
\newcommand{\pgao}[1]{\textcolor{cyan}{-PG: #1-}}

\newcommand{\tool}{\textsc{CHAMP}\xspace}
\newcommand{\sys}{\textsc{CHAMP}\xspace}

\newcolumntype{L}[1]{>{\raggedright\let\newline\\\arraybackslash\hspace{0pt}}m{#1}}
\newcolumntype{C}[1]{>{\centering\let\newline\\\arraybackslash\hspace{0pt}}m{#1}}
\newcolumntype{R}[1]{>{\raggedleft\let\newline\\\arraybackslash\hspace{0pt}}m{#1}}

\def\BibTeX{{\rm B\kern-.05em{\sc i\kern-.025em b}\kern-.08em
    T\kern-.1667em\lower.7ex\hbox{E}\kern-.125emX}}

\begin{document}

\title{\tool: Characterizing Undesired App Behaviors from User Comments based on Market Policies}

\author{
\IEEEauthorblockN{Yangyu Hu$^{1*}$, Haoyu Wang$^{2*}$\Letter
\thanks{*The first two authors contributed equally to this work. Prof. Haoyu Wang is the corresponding author (haoyuwang@bupt.edu.cn).}, Tiantong Ji$^{3}$, Xusheng Xiao$^{3}$, Xiapu Luo$^{4}$, Peng Gao$^{5}$ and Yao Guo$^{6}$}
\IEEEauthorblockA{
$^{1}$ Chongqing University of Posts and Telecommunications, Chongqing, China\\
$^{2}$ Beijing University of Posts and Telecommunications, Beijing, China\\	
$^{3}$ Case Western Reserve University, USA	
$^{4}$ The Hong Kong Polytechnic University, Hong Kong, China\\
$^{5}$ University of California, Berkeley, USA
$^{6}$ Peking University, Beijing, China\\
}
}
\maketitle
\begin{abstract}
Millions of mobile apps have been available through various app markets. Although most app markets have enforced a number of automated or even manual mechanisms to vet each app before it is released to the market, thousands of low-quality apps still exist in different markets, some of which violate the explicitly specified market policies. 
In order to identify these violations accurately and timely, we resort to user comments, which can form an immediate feedback for app market maintainers, to identify undesired behaviors that violate market policies, including security-related user concerns.
Specifically, we present the first large-scale study to detect and characterize the correlations between user comments and market policies.
First, we propose \tool, an approach that adopts text mining and natural language processing (NLP) techniques to extract semantic rules through a semi-automated process, and classifies comments into 26 pre-defined types of undesired behaviors that violate market policies. Our evaluation on real-world user comments shows that it achieves both high precision and recall ($>0.9$) in classifying comments for undesired behaviors.
Then, we curate a large-scale comment dataset (over 3 million user comments) from apps in Google Play and 8 popular alternative Android app markets, and apply \tool to understand the characteristics of undesired behavior comments in the wild. The results confirm our speculation that user comments can be used to pinpoint suspicious apps that violate policies declared by app markets. The study also reveals that policy violations are widespread in many app markets despite their extensive vetting efforts. \tool can be a \textit{whistle blower} that assigns policy-violation scores and identifies most informative comments for apps.
\end{abstract}

\begin{IEEEkeywords}
User comment, app market, undesired behavior
\end{IEEEkeywords}

\section{Introduction}

Although the mobile app ecosystem has seen explosive growth in recent years, app quality remains a major issue across app markets~\cite{wang2018beyond,wang2019understanding}.
On the one hand, it is reported that millions of Android malicious apps were identified every year~\cite{malware}, using more and more complex and sophisticated malicious payloads and evasion techniques~\cite{tam2017evolution,tang2020all,hu2018dating}.
On the other hand, a large number of fraudulent and gray behaviors (e.g., ad fraud) were found in the mobile app ecosystem from time to time~\cite{xi2019deepintent,appsquatting,dong2018frauddroid, andow2016study,liu2020maddroid,liu2019dapanda}.
Furthermore, apps with functionality/performance issues such as ``\textit{diehard apps}''~\cite{zhou2020demystifying}, and devious contents such as ``anti-society contents'' still remain in the markets~\cite{wang2018android}.

Most app markets have released strict developer policies, along with inspection and vetting processes before app publishing, seeking to nip the aforementioned threats in the bud and improve app quality in the markets.
For example, Google Play has released a set of developer policies~\cite{gplaypolicy} that cover 10 main categories, including ``Privacy, Security and Deception'', ``Spam and Minimum Functionality'', and ``Monetization and Ads'', etc.
Each category stands for a type of violation that may be associated with various undesired behaviors.
Apps that break these policies should not be published on Google Play.

However, it is challenging to automatically check policy compliance for mobile apps.
Despite Google Play's efforts in adopting strict vetting processes by using automated tools~\cite{bouncer, googleAI}, malware and Potentially Harmful Apps (PHAs) are recurrently found in Google Play~\cite{wang2019rmvdroid}.
Third-party app markets also show a significantly higher prevalence of malware, fake, and cloned apps~\cite{wang2018beyond}.
On the one hand, it has been reported that many malicious apps use sophisticated techniques to evade automated detection~\cite{tam2017evolution}. For example, certain malicious behaviors could only be triggered at a specific time or environment, such as checking whether the app is being inspected in emulating environments~\cite{petsas2014rage}.
On the other hand, even if malware can be detected by these automated tools, many other fraudulent and gray behaviors such as ad fraud and malicious push notifications are hard to identify.
Moreover, functionality/performance issues are typically \emph{app-specific},
while devious contents are broad and difficult to detect without human inspection, posing more challenges for automated tools~\cite{wang2018android}.

\begin{figure}[t]
\centering
  \includegraphics[width=0.49\textwidth]{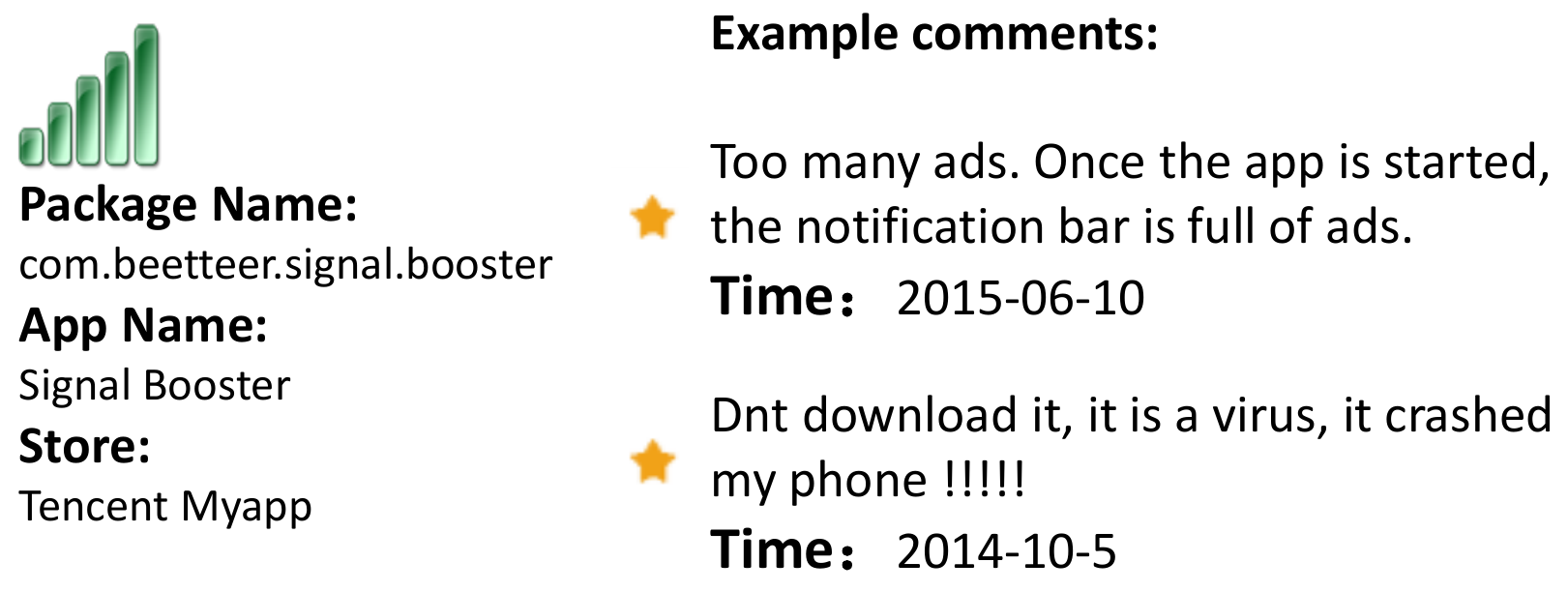}
\caption{An example of user-perceived undesired behavior.}
\label{fig:example}
\end{figure}

In many cases, whether an app's behavior has exposed severe security risks or performance issues depends on how users think of it~\cite{autoreb,lovereview}.
As an important process for developers to improve app quality, app markets allow users to leave their ratings and comments after downloading and using each app~\cite{appreview}.
These comments can be considered as the direct feedback from users who have experienced the apps~\cite{lovereview}, helping developers address the issues that might not have been spotted in testing.
For example, as shown in Figure~\ref{fig:example}, two users gave 1-star ratings for the app.
One user complained that this app contains aggressive advertising behaviors, and the other even reported that this app might be malicious.
In fact, this behavior is also one of the \emph{undesired behaviors} explicitly prohibited by the developer policies.
When such comments are made aware to the market maintainers, they should be able to warn the app developers about the behaviors immediately and remove the apps from the market if such undesired behaviors are not addressed by the app developers.
In other words, \emph{user comments can form an immediate feedback for app market maintainers to identify user concerns and characterize the undesired behaviors that violate market policies.}

Ideally, user reviews could serve as an effective source for app markets to identify policy violations in the apps after they have passed the initial vetting process. 
However, the number of user comments in an app market is huge given the rapidly increasing number of apps, and there is a lack of automated tools to detect comments that are related to market policy violations and further perform deeper analysis on these comments.
Furthermore, these useful comments are often buried in a much larger number of irrelevant comments, making it labor-intensive and error-prone to manually inspect these comments to obtain feedback.
While user comments have been studied for emerging issues~\cite{reviewbug}, app risks~\cite{autoreb} and app recommendation~\cite{LiReview},
few research efforts have been spent in investigating how user comments can assist app markets in improving app vetting process.
Thus, little is known \emph{to what extent user comments can provide feedback on undesired behaviors that violate market policies and how app markets can utilize these feedback to improve their app vetting and maintenance process.}

In this work, we investigate the correlation between user comments and market policies, \ie characterizing user-perceived undesired behaviors prohibited by market policies.
First, we create a taxonomy of 26 kinds of undesired behaviors summarized from the developer policies of 9 app markets.
Then, we propose \tool, an approach that adopts text mining and NLP techniques to identify comments that describe these 26 kinds of undesired behaviors and classify them.
We refer to such comments as \emph{\textbf{undesired-behavior comments (\textit{UBComments})}}.
More specifically, \tool first extracts semantic rules from a training dataset of user comments via a semi-automated process.
\tool then uses the extracted rules to automatically identify the undesired behaviors reflected in a given comment.
Evaluation of \tool on benchmarks from real-word user comments suggests that it can successfully identify \textit{UBComments} with high precision and recall ($>$0.9).

To further understand \textit{UBComments} in the wild, we have curated a large-scale dataset from 9 app markets, with over 3 million user comments. We applied \tool on these comments to identify the \textit{UBComments} and study their characteristics. We have a number of interesting findings:

\begin{itemize}
    \item \textit{UBComments} are prevalent in the app ecosystem, which can be found in 47\% of the apps we studied.
    UBComments account for 20\% for the 1-star comments. Our manual verification on sampled apps suggested the existence of undesired behaviors (96\% of them could be verified).
    \textbf{It confirms our assumption that users can still perceive a large number of undesired behaviors prohibited by market policies, even though these apps have already passed the comprehensive vetting process}. 

    \item User-perceived undesired behaviors, even some security-related ones, can be found in both malware and ``benign'' apps (the apps that were not flagged by any anti-virus engines on VirusTotal~\cite{virustotal}). \textbf{It suggests that user comments can be a complementary source for providing insights of malware detection.}
    
    \item Although each market has explicitly declared developer policies, roughly 34\% to 65\% of apps in each market were still complained about their undesired behaviors against the policies. \textbf{This observation further indicates that it is hard for app markets to identify all policy violations during app vetting, while user comments could further help detect these violations continuously.}
    Moreover, policies from most markets are inadequate, as we have identified many apps (5\% to 60\%) showing undesired behaviors that are not covered in their policies.

\end{itemize}

To the best of our knowledge, this is the first large-scale study on the \emph{correlation between user comments and market policies of mobile apps}. 
We believe that our research efforts can positively contribute to the app vetting process, promote best operational practices across app markets, and boost the focus on related topics for the research community and market maintainers.
We have released the \tool tool, along with the policies and dataset to the research community at Github~\cite{UBCFinderGithub}.

\begin{table}[t]
\caption{The distribution of policies collected (total 599).}
\label{table:marketpolicydistribution}
\centering
\resizebox{\linewidth}{!}{
\begin{tabular}{lc||lc}
\toprule
Market &  \# Policies & Market &  \# Policies\\ 
\midrule
GooglePlay~\cite{googleplay} & 172 & 360 Market~\cite{360market} & 30\\ 
Huawei Market~\cite{huawei} & 22 & Lenovo Market~\cite{lenovo} & 28\\
Meizu Market~\cite{meizu} & 53 & Oppo Market~\cite{oppo} & 15\\
Vivo Market~\cite{vivo} & 96 & Xiaomi Market~\cite{xiaomi} & 159\\
Tencent Myapp~\cite{tencentmyapp} & 24 & &\\
\bottomrule
\end{tabular}
}
\end{table}

\begin{table*}[t]
\caption{A taxonomy of undesired behaviors and the distribution across market policies. The \Checkmark refers to the market declaring the policies. The number refers to the \# of apps with \textit{UBcomments} we identified from each market in Section~\ref{sec:measurement}.}
\label{table:undesiredbehavior}
\centering
\resizebox{\linewidth}{!}{
\begin{tabular}{c|c|ccccccccc}
\toprule
Category & Behavior & \tabincell{c}{360 \\Market} & Huawei & Lenovo & Meizu & Oppo & Vivo & Xiaomi & \tabincell{c}{Tencent \\ Myapp} & \tabincell{c}{Google \\ Play}\\ 
\midrule
\multirow{8}*{\tabincell{l}{Functionality and \\ Performance}} & fail to install & 264 & \Checkmark (\textbf{216}) & \Checkmark (\textbf{70}) & \Checkmark (\textbf{50}) & 33 & \Checkmark (\textbf{32}) & \Checkmark (\textbf{39}) & \Checkmark (\textbf{106}) & \Checkmark (\textbf{5})\\
~ & fail to retrieve content & 30 & 33 & 5 & \Checkmark (\textbf{9}) & 10 & \Checkmark (\textbf{11}) & \Checkmark (\textbf{12}) & \Checkmark (\textbf{11}) & \Checkmark (\textbf{21})\\
~ & fail to uninstall & \Checkmark (\textbf{119}) & \Checkmark (\textbf{46}) & \Checkmark (\textbf{11}) & \Checkmark (\textbf{19}) & 23 & \Checkmark (\textbf{29}) & \Checkmark (\textbf{21}) & \Checkmark (\textbf{49}) & \Checkmark (\textbf{1})\\
~ & fail to start (e.g., crash) & 699 & \Checkmark (\textbf{451}) & \Checkmark (\textbf{209}) & \Checkmark (\textbf{238}) & \Checkmark (\textbf{174}) & \Checkmark (\textbf{318}) & 176 & \Checkmark (\textbf{880}) & \Checkmark (\textbf{105})\\
~ & bad performance (e.g., no responding) & 334 & \Checkmark (\textbf{134}) & 30 & 60 & \Checkmark (\textbf{53}) & \Checkmark (\textbf{65}) & \Checkmark (\textbf{41}) & 176 & \Checkmark (\textbf{18})\\
~ & fail to login or register & 180 & 201 & \Checkmark (\textbf{52}) & 88 & 98 & \Checkmark (\textbf{143}) & 86 & 184 & \Checkmark (\textbf{33})\\
~ & fail to exit & \Checkmark (\textbf{62}) & 45 & 4 & 11 & 11 & 10 & 9 & 15 & \Checkmark (\textbf{2})\\
~ & powerboot & \Checkmark (\textbf{3}) & 1 & 1 & \Checkmark (\textbf{0}) & 0 & \Checkmark (\textbf{0}) & \Checkmark (\textbf{3}) & \Checkmark (\textbf{0}) & \Checkmark (\textbf{5})\\
\midrule
\multirow{4}*{\tabincell{l}{Advertisement}} & drive-by download & 25 & 22 & 5 & 13 & 7 & 6 & \Checkmark (\textbf{14}) & \Checkmark (\textbf{9}) & \Checkmark (\textbf{25})\\
~ & ad disruption & \Checkmark (\textbf{498}) & 262 & \Checkmark (\textbf{91}) & \Checkmark (\textbf{180}) & \Checkmark (\textbf{118}) & \Checkmark (\textbf{168}) & \Checkmark (\textbf{145}) & \Checkmark (\textbf{818}) & \Checkmark (\textbf{167})\\
~ & add shortcuts in launching menu & 7 & 1 & 0 & \Checkmark (\textbf{7}) & 1 & \Checkmark (\textbf{0}) & 4 & \Checkmark (\textbf{4}) & \Checkmark (\textbf{7})\\
~ & ads in notification bar & 15 & 1 & \Checkmark (\textbf{0}) &\Checkmark (\textbf{3})  & 1 & \Checkmark (\textbf{1}) & \Checkmark (\textbf{1}) & \Checkmark (\textbf{10}) & \Checkmark (\textbf{2})\\
\midrule
\multirow{10}*{\tabincell{l}{Security}} & virus & \Checkmark (\textbf{139}) & \Checkmark (\textbf{96}) & \Checkmark (\textbf{18}) & \Checkmark (\textbf{39}) & \Checkmark (\textbf{40}) & \Checkmark (\textbf{45}) & \Checkmark (\textbf{33}) & \Checkmark (\textbf{151}) & \Checkmark (\textbf{54}) \\
~ & privacy leak & \Checkmark (\textbf{25}) & 24 & 5 & 7 & \Checkmark (\textbf{9}) & \Checkmark (\textbf{16}) & \Checkmark (\textbf{11}) & 24 & \Checkmark (\textbf{30})\\
~ & payment deception & \Checkmark (\textbf{236}) & \Checkmark (\textbf{189}) & \Checkmark (\textbf{39}) & 74 & 84 & \Checkmark (\textbf{127}) & \Checkmark (\textbf{61}) & 282 & \Checkmark (\textbf{75})\\
~ & \tabincell{c}{illegal background behavior (e.g., sms)} & 160 & 109 & 24 & \textbf{57} & 51 & \Checkmark (\textbf{49}) & \Checkmark (\textbf{44}) & \Checkmark (\textbf{146}) & \Checkmark (\textbf{0})\\
~ & excessive network traffic & \Checkmark (\textbf{90}) & 40 & 3 & 13 & \Checkmark (\textbf{25}) & \Checkmark (\textbf{30}) & \Checkmark (\textbf{16}) & 111 & \Checkmark (\textbf{4})\\
~ & hidden app & \Checkmark (\textbf{12}) & 1 & 2 & 4 & \Checkmark (\textbf{0}) & \Checkmark (\textbf{0}) & \Checkmark (\textbf{1}) & 2 & \Checkmark (\textbf{1})\\
~ & illegal redirection & 80 & 35 & \Checkmark (\textbf{5}) & 17 & 20 & \Checkmark (\textbf{19}) & \Checkmark (\textbf{16}) & 135 & \Checkmark (\textbf{8})\\
~ & \tabincell{c}{permission abuse} & 37 & \Checkmark (\textbf{27}) & 4 & \Checkmark (\textbf{8}) & 4 & \Checkmark (\textbf{4}) & \Checkmark (\textbf{17}) & \Checkmark (\textbf{11}) & \Checkmark (\textbf{27}) \\
~ & illegitimate update (e.g., update to other app) & 3 & 3 & \Checkmark (\textbf{0}) & 0 & 3 & 1 & 2 & 1 & \Checkmark (\textbf{0})\\
~ & browser setting alteration & 0 & 0 & 0 & 0 & 0 & \Checkmark (\textbf{0}) & \Checkmark (\textbf{0}) & 0 & \Checkmark (\textbf{0})\\
\midrule
\multirow{2}*{\tabincell{c}{Illegitimate Behavior \\of Developers}} & app repackaging & 132 & 16 & 12 & \Checkmark (\textbf{11}) & 14 & 17 & \Checkmark (\textbf{13}) & 64 & \Checkmark (\textbf{14})\\
~ & app ranking fraud & 54 & 28 & 7 & \Checkmark (\textbf{34}) & 22 & \Checkmark (\textbf{20}) & \Checkmark (\textbf{21}) & 45 & \Checkmark (\textbf{6}) \\
\midrule
\multirow{2}*{\tabincell{l}{Content}} & vulgar content (e.g., pornography, anti-society) & \Checkmark (\textbf{47}) & 18 & \Checkmark (\textbf{1}) & \Checkmark (\textbf{6}) & 4 & \Checkmark (\textbf{8}) & 14 & \Checkmark (\textbf{21}) & \Checkmark (\textbf{15})\\
~ & inconsistency between functionality and description & 15 & 5 & \Checkmark (\textbf{0}) & 2 & 3 & 8 & \Checkmark (\textbf{1}) & 8 & \Checkmark (\textbf{1})\\
\midrule
\multicolumn{2}{c|}{Total \# of apps with undesired behaviors} & 1025 & 625 & 338 & 422 & 237 & 463 & 257 & 1382 & 274\\
\midrule
\multicolumn{2}{c|}{Total \# of apps with undesired behaviors (declared policies)} & 731 & 537 & 318 & 365 & 210 & 460 & 211 & 1233 & 274\\
\midrule
\multicolumn{2}{c|}{Total \# of apps with undesired behaviors (undeclared policies)} & 891 & 433 & 90 & 219 & 178 & 36 & 191 & 654 & 0\\
\bottomrule
\end{tabular}
}
\end{table*}

\section{A Taxonomy of Undesired Behaviors}
\label{sec:background}
\label{sec:undesiredbehavior}

As we seek to identify the \textit{UBComments} and investigate the correlation between user comments and market policies, we first collect a dataset of market policies and compile a taxonomy of the undesired behaviors described in them.

\textbf{Market Policy Dataset.}
Considering that Google Play is the dominating market in the world except China, we seek to collect policies from 9 popular markets, including Google Play and 8 top Chinese third-party app markets, as shown in Table~\ref{table:marketpolicydistribution}. For each market, we crawl all the listed policies from the corresponding webpages. In total, we have collected 599 policies.
Note that the developer policies of Google Play were in English, while the other market policies were in Chinese. 
Google Play has more complete and fine-grained policies than any of the third-party app markets.

\textbf{Summary of Undesired Behaviors.}
As the policies defined by each market vary greatly (some are coarse-grained and some are fine-grained), it is non-trivial to automatically classify them.
Thus, the first two authors of this paper manually went through these policies, and classified them into 5 main categories, including 26 distinct undesired behaviors.
Table \ref{table:undesiredbehavior} shows the taxonomy of the summarized undesired behaviors, and the distribution of the corresponding policies across markets. 
Note that one behavior may correspond to one or more market policies.
We observe that all of the undesired behavior regulations can be found in Google Play. 
As for the third-party markets, \texttt{Vivo} and \texttt{Xiaomi} have declared policies related to the most types of undesired behaviors, covering 21 and 20 behaviors respectively. 
We believe that this taxonomy covers most of the commonly observed undesired behaviors.
Even though it may still be incomplete, our approach is generic and can be adapted to support new behaviors and different granularities of behaviors (see Section~\ref{sec:discussion}).

\section{Automated Classification of UBComments}
\label{sec:behaviorextraction}

\begin{figure*}[t]
\centering
  \includegraphics[width=0.99\textwidth]{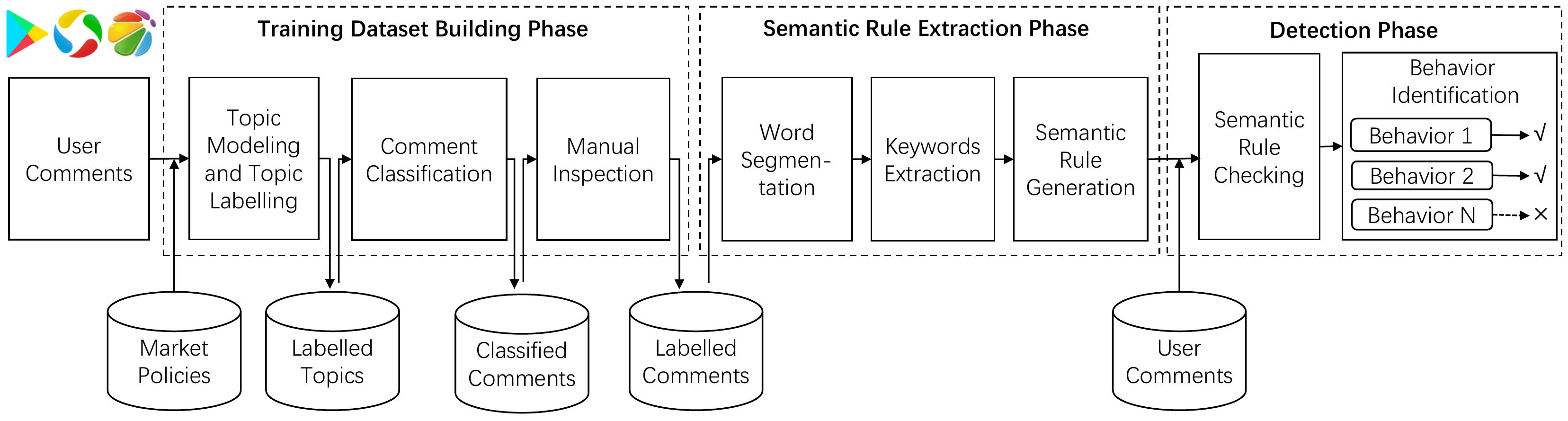}
\caption{Overview of \tool.}
\label{fig:method}       
\end{figure*}

\subsection{Overview}
\label{sec:methodlogy}

Figure~\ref{fig:method} shows the overview of \tool, which builds a training dataset of user comments (the \emph{training dataset building phase}), extracts semantic rules from the labelled comments (the \emph{semantic rule extraction phase}) and uses the rules to identify and classify \textit{UBComments} (the \emph{detection phase}). 
The major reason why we prefer semantic rules instead of text similarity is that most comments are short and often use a few key phrases in specific orders such as ``icon disappears'',
while semantic rules have shown promising results in identifying sentences with specific purposes~\cite{xiao12:text2policy,pandita12:inferring,pandita2013whyper}.
On the contrary, text similarity approaches based on word similarity without emphasis on key phrases are optimized for general purposes, 
and thus these approaches require extra tuning to focus on certain words that play important roles in the sentences of market policies~\cite{textsimilarity1,li2006sentence,islam2008semantic}. Additionally, these approaches generally require a substantial amount of labelled samples to train the weights, which is less effective in our context due to the limited number of labelled samples. 

\ding{182}
In the training dataset labelling phase, we collect the comments of the apps from Google Play and 8 third-party app markets, and resort to text clustering model to help to label the user comments.
In the topic modeling and topic labelling step, we first merge the market policies that describe a same undesired behavior into a single document (26 documents in total). Then, \tool applies a short-text topic modeling algorithm~\cite{BTM,BTMcite1,BTMcite2,BTMcite3} to identify a set of topics, where each topic contains a set of words.
At last, \tool labels each topic with related undesired behavior based on the similarity between the documents of policies and the words in the topics.
In the comment classification step, \tool uses the labelled topics to classify each comment into related undesired behaviors.
We further manually inspect the classified comments to confirm whether these comments are related to the corresponding undesired behavior. This is necessary because we could only classify each comment based on the keywords with the highest weight under each topic, which may introduce false positives. For example, if a comment contains the keyword ``notification'', it is considered to be likely to related to the behavior ``ads in notification bar". However, the word ``notification'' may also appear in comments that talk about alerts and notifications (e.g., notifications and alerts for weather apps).
\ding{183}
In the semantic rule extraction phase, \tool applies a generation algorithm on the labelled comments and generates semantic rules for each undesired behavior automatically.\ding{184}
In the detection phase, \tool accepts user comments as input, and uses the semantic rules to classify comments into the undesired behaviors defined in market policies.

\subsection{Training Dataset Labelling}
\label{sec:trainingdataset}

\textbf{Training Dataset.}
To label training dataset, we randomly select 2\% of the comments for each app in our dataset (discussed in \S~\ref{sec:dataset}). In total, we extract 70,000 comments, including 15,000 English comments and 55,000 Chinese comments. 
Note that these comments were used separately for training two models for both English and Chinese comments.

\textbf{Topic Modeling and Topic Labelling}.
Unlike traditional documents (\eg news articles), the descriptions of undesired behaviors in market policies consist of only one or a few short sentences. Thus, the lack of rich context makes it infeasible to use the topic modeling algorithms such as PLSA~\cite{plsa} and LDA~\cite{lda}, which implicitly model document-level word co-occurrence patterns. To address this problem, we apply BTM (biterm topic model)~\cite{BTM}, a widely used model for short-text topic modeling, to learn the set of topics for market policies. 
BTM explicitly models word co-occurrence patterns using \emph{biterms}, where each biterm is an unordered word-pair co-occurred in a short context.
The output of BTM are a set of topics where each topic consists of a list of words and their weights.
For each topic $z$, BTM draws a topic-specific word distribution $\phi_{z} \sim Dir(\beta)$, and
draws a topic distribution $\theta \sim Dir(\alpha)$ for all of the documents, where $\alpha$ and $\beta$ are the Dirichlet priors. For each biterm $b$ in the biterm set $B$, it draws a topic assignment $Z \sim Multi(\theta)$ and draws two words $(w_{i}, w_{j}) \sim Multi(\phi_{z})$, where $w_{i}$ and $w_{j}$ are words appearing in the same document. 
Following the above procedure, the joint probability of a biterm $b = (w_{i}, w_{j})$ can be written as:
$$P(b) = \sum\limits_{z}P(z)P(w_{i}|Z)P(w_{j}|Z)$$
 thus the likelihood of all the documents is:
$$P(B) = \quad\prod_{(i,j)}\sum\limits_{z}\theta(z)\phi_{i|z}\phi_{j|z}$$

We conduct topic modeling based on the merged English and Chinese policies, respectively. We set the number of topics as 26, which corresponds to the number of undesired behaviors.
\tool then labels the proper undesired behaviors for the topics by computing the probability of each document being allocated to each topic. It assumes that the topic proportion of a document equals to the expectation of the topic proportion of generated biterms during topic modelling:
$$P(z|d) = \sum\limits_{b}P(z|b)P(b|d),$$
where $z$ represents topic, $b$ represents biterm and $d$ represents document. $p(z|b)$ can be calculated via Bayes formula based on the parameters estimated in BTM:
$$P(z|b) = \frac{\theta_{z}\phi_{i|z}\phi_{j|z}}{\sum\limits_{z}\theta_{z}\phi_{i|z}\phi_{j|z}}$$
$p(b|d)$ can be estimated by the empirical distribution of biterms in the document, where $n_{d}(b)$ is the frequency of the biterm $b$ in the document $d$:
$$p(b|d) = \frac{n_{d}(b)}{\sum\limits_{b}n_{d}(b)}$$

At last, \tool selects the highest score of P(z|d) and labels the proper undesired behaviors for each of the 26 topics.

\textbf{Comment Classification.}
\tool then classifies each comment into related topics. 
It computes the probability of each comment being allocated to each topic.
If the probability is above a certain threshold, \tool considers that the comment is related to the topic.
We follow the same empirical approach~\cite{BTMcite1, BTMcite2, BTMcite3} to set the threshold, and find that 0.6 is a good indicator. In total, we obtain 9,228 comments that are related to 26 distinct behaviors.

\textbf{Manual Inspection.}
Considering that the automated classified comments may be not related to the undesired behaviors (see \S~\ref{sec:behaviorextraction}.A ), we further manually inspected the comments that are classified into related topics to confirm whether these comments are \textit{UBComments}. Besides, if a comment is related to more than one behavior, we split the comment into several sentences and each sentence is related to a kind of undesired behavior. Two authors inspect the comments independently. For the disagreements of category labelling, a further discussion is performed.
Eventually, we obtained 8,275 comments that are related to 25 distinct behaviors. After splitting some comments, we obtained 9,057 labelled comments in total, which will be used for semantic rules generation. Note that we did not find any comments that are related to the behavior of ``browser setting alteration''.

\begin{table}[t]
\caption{Representative stopwords used in \tool.}
\label{table:stopwords}
\centering
\begin{tabular}{c|c}
\toprule
\tabincell{c}{Removed \\Stopwords} & \tabincell{c}{Added \\stopwords} \\ 
\midrule
\tabincell{c}{miss, high, ask, give, \\can not, how, able, stop, \\without, allow, obtain, other} & \tabincell{c}{god, sex, s**t, s**d, \\silly, blah, r**h, d**n, \\d**b, da*n, horrible} \\
\bottomrule
\end{tabular}
\end{table}

\subsection{Automated Semantic Rule Extraction}
\label{sec:semanticrule} 
Based on the labelled comments, given a new comment, the goal of \tool is to determine whether the comment describes the same or similar behavior as the labelled comments. To achieve this goal, we propose to automatically extract \textit{semantic rules} from the labelled comments for each undesired behavior. Firstly,  for each undesired behavior, \tool extracts and sorts the representative words from the related comments. Then, \tool analyzes the relations of the keywords by merging the keywords that usually appear in the same comments.
After that, we can get one or more keyword sets containing different representative keywords. At last, \tool generates semantic rules for each keyword set by combining keywords and calculating the distance constraints of the keywords.

\textbf{Word Segmentation.}
In this step, \tool groups the comments related to each undesired behavior into a corpus (25 corpora in total). For each corpus, it segments the comments into words, removes meaningless words and sorts the remaining words in descending order based on the TF-IDF~\cite{tfidf} weighting to generate a word list ${WordList}$.
Stopwords are the words considered
unimportant in text analysis tasks.
Thus, we take advantage of the stopword lists provided by HIT~\cite{stopwords} and a public English stopwords list ``stopwords-iso''~\cite{stopwordsen}. \textit{However, we find that the general stopword lists cannot well fit the app comment study}. 
On one hand, when some traditional stopwords (\eg can) are combined with other words, they become key phrases for describing undesired behaviors in user comments. For example, the comment ``always have to download other apps'' is related to the undesired behavior ``drive-by download'', thus the traditional stopwords ``always'' and ``other'' should not be removed. 
We summarized and removed 29 stopwords (including 14 English stopwords and 15 Chinese stopwords) that are important for describing undesired behaviors from the stopwords list. On the other hand, existing research found that there exist
noises and spams (e.g., offensive comments) in app comments, which are meaningless for describing undesired behaviors. Therefore, we adapt the selected stopword list and add over 50 new stopwords that are regularly appeared in user comments. 
The representative stopwords are shown in Table~\ref{table:stopwords} (offensive words are sanitized).

\textbf{Representative Keywords Extraction.}
The goal of this step is to identify the most representative keywords that can cover the labelled comments in a given corpus. Thus, for each keyword in the \textit{WordList} of a given corpus, \tool first collects the comments in the corpus that contain the keyword and adds them into a comment set ${ComtSet}_{word}$. Then, a traversal operation begins to select the keywords in order (based on TF-IDF weight) and compare the ${ComtSet}_{word}$ of different words. For the comment set ${ComtSet}_{word_m}$ of the m-th word ${word}_{m}$ in the ${WordList}$, if part of the comments in it are overlapped with the comments in the n-th word's ($n < m$) comment set ${ComtSet}_{word_n}$, \tool will merge ${word}_{m}$ and ${word}_{n}$ into a keyword set. Otherwise, \tool will assign the word ${word}_{m}$ into a new keyword set. 
Note that, the traversal operation will stop if the union set from ${ComtSet}_{word_1}$ to ${ComtSet}_{word_m}$ contains all of the labelled comments in the corpus. Based on the traversal operation, \tool could extract one or more keyword sets for each corpus.

\textbf{Semantic Rule Generation.}
For each of the extracted keyword sets in a corpus, \tool automatically generates semantic rules. We observe that a behavior can be generally described by two keywords of different part-of-speech~\cite{partofspeech} in a comment. For example, the verb ``steal'' and the noun ``money'' in the comment ``it steals money from the credit card!!!'' are related to behavior ``payment deception''. Another example, the adverb ``how'' and the verb ``uninstall'' in the comment ``who can tell me how to uninstall this app'' are related to behavior ``fail to uninstall''. Thus, for the extracted keyword sets, \tool combines the keywords of different part-of-speech pairwise. Furthermore, we observe that most \textit{UBComments} are short and often include key phrases in specific orders. Therefore, the semantic rules not only contain keywords but include order and distance constraints on matching the keywords. For two keywords ${keyword}_{u}$ and ${keyword}_{v}$ $({ComtSet}_{u}\cap{ComtSet}_{u}\neq\emptyset)$, \tool will generate two semantic rules $\{{keyword}_{u},{keyword}_{v},constraints\}$ and $\{{keyword}_{v},{keyword}_{u},constraints\}$, the constraints is used to limit the distances of these two keywords. For example, semantic rule $\{ask,permission,3\}$ means that ``ask'' appears before ``permission'' and their distance is less than 3 words. \tool automatically calculates the F1-score under different distance constraints (we set it range from 1 to 20) for each semantic rule, and select the best one. Note that, if all of the keywords in a keyword set are noun, each keyword will generate a semantic rule $\{{keyword},null,null\}$. 

Eventually, 
\tool generates 320 semantic rules for the 26 undesired behaviors in total (the list of rules can be found in~\cite{UBCFinderGithub}), in which 136 semantic rules are for English comments and 184 semantic rules are for Chinese comments. 
Note that there are no comments related to the behavior of ``modify browser setttings'' and thus we use the description in the related policies to extract semantic rules (4 rules in total).
The major differences between Chinese comment rules and English comment rules are \emph{synonyms}. Synonyms in Chinese are more frequently used than in English, leading to more rules for some undesired behaviors. For example, two keywords ``uninstall'' and ``remove'' of the semantic rules for behavior ``fail to uninstall'' are generated in English comments, while \tool has extracted 5 synonyms of these two keywords in Chinese comments. 
Table~\ref{table:heuristicrule} shows representative semantic rules for 3 undesired behaviors in English comments (the complete set of rules can be found at Github~\cite{UBCFinderGithub}). As our semantic rules are trained to detect similar sentences that describe the behaviors in the policies, thus the detected sentences are all high quality, which will be evaluated in \S~\ref{sec:experiment}.

\begin{table}[t]
\caption{Representative semantic rules for 4 behaviors.}
\label{table:heuristicrule}
\centering
\resizebox{0.8\linewidth}{!}{
\begin{tabular}{c|c}
\toprule

Behavior & semantic rules \\
\midrule
\multirow{3}*{\tabincell{c}{virus}} & \{virus, null, null\}\\
~ & \{trojan, null, null\}\\
~ & \{malware, null, null\}\\
\midrule
\multirow{4}*{\tabincell{c}{ads in notification bar}} & \{notification, ads, 3\}\\
~ & \{notification, full, 2\}\\
~ & \{remove, notification, 4\}\\
\midrule
\multirow{5}*{\tabincell{c}{permission abuse}} & \{ask, permission, 5\}\\
& \{require, permission, 6\}\\
& \{unnecessary, permission, 2\}\\
& \{need, permission, 6\}\\
& \{want, permission, 7\}\\
\bottomrule
\end{tabular}
}
\end{table}

\subsection{Semantic Rule Checking}
Based on these semantic rules, \tool classifies each comment into a type of \textit{UBComments} or others. 
Given a comment, \tool first removes the stopwords and performs word segmentation~\cite{jieba} to extract words from the comment. 
\tool then applies the semantic rules one by one to determine whether the comment matches any rules.
It searches the extracted words to see whether the keywords appear in the extracted words and checks the order and distance of successful matching keywords to determine whether they meet the constraints of the semantic rules.
As shown in Fig.~\ref{fig:example}, the motivating app violates two behaviors, i.e., ``ads in notification bar'' and ``virus''.
Based on the rules defined in Table~\ref{table:heuristicrule}, \tool determines that the first comment ``too many ads, ..., the notification bar is full of ads'' matches 2 semantic rules of the undesired behavior ``notification bar'', since the comment has the keywords of ``notification'' , ``ads'' and ``full''. Similarly, the other comment contains the keyword of ``virus'' and thus matches the undesired behavior ``virus''.
\section{Study Design}
\label{sec:dataset}
\eat{
The objective of our study is (1) to answer to what extent we can infer the undesired behaviors from user comments, (2) measure the regulation behaviors across markets based on user comments and characterize the reactions to such comments, and (3) provide insights to app market maintainers on how to apply such techniques to better identify apps with undesired behaviors.
To this end, we first summarize a list of key research questions, and then provide how we collect a diverse and large-scale dataset for evaluation.
}

\subsection{Research Questions}
We seek to answer the following research questions (RQs):

\begin{itemize}
    \item[RQ1] \textbf{How effective is \tool in detecting \textit{UBComments}?} 
    As we aim to apply \tool to extract \textit{UBComments} in the wild, 
    It is necessary to first evaluate the effectiveness of \tool on extracting undesired behaviors using a benchmark dataset. 
    
    \item[RQ2] \textbf{What kinds of undesired behaviors can be perceived by users?} 
    It is important to explore to what extent we can infer undesired behaviors from user comments, and which behaviors can be perceived by users. 
    
    \item[RQ3] \textbf{How well do the policies in each app market capture the undesired behaviors reflected by user comments?} 
    As each app market has its own policies, we want to know whether they are effective in flagging undesired behaviors during the app vetting process.
    App markets with weak app vetting processes are more likely to be exploited.
    
\end{itemize}

\subsection{Dataset}

\subsubsection{Collecting App Candidates}

To answer the RQs, we first need to harvest a comprehensive dataset that covers as many undesired behaviors as possible. 
We take advantage of existing efforts, and use a large-scale Android app repository~\cite{wang2018beyond}.
This repository contains over 6.2 million app items collected from Google Play and 17 third-party app markets. 
The dataset also provides the detection result of VirusTotal~\cite{virustotal}, a malware analysis service that aggregates over 60 anti-virus (AV) engines.
To better understand the distribution of \textit{UBComments} across apps with different maliciousness levels, we classified our app candidates into 3 categories: malware, grayware and benign apps. 
As previous studies~\cite{wei2017deep} suggested that some AV engines may not always report reliable results, we regard the apps labeled by over half of the AV engines (>30) as malware, which is supposed to be a reliable threshold by previous work~\cite{wei2017deep}. 
We consider apps flagged by no AV engines as benign apps, and the other apps as grayware. 
This roughly classification of malware and grayware might not be accurate enough, but this is not the focus of this paper. As the number of reported engines can be used as an indicator of the maliciousness of the apps, we only want to study the diversity across apps with different levels of maliciousness.
We randomly selected 10,000 target app candidates (8,400 Chinese apps and 1,600 Google Play apps) from the dataset of Wang et al.~\cite{wang2018beyond}, including 4,000 malware, 3,000 grayware and 3,000 benign apps. Note that the 1,600 Google Play apps include 1,000 benign apps and 600 grayware, as all the malware samples were removed by Google Play and we cannot get their comments (NA in Table~\ref{table:comment}).

\subsubsection{Harvesting the User Comments}
\label{sec:comment}

All the app markets we studied only provide a limited number of user comments. For example, Google Play review collection service~\cite{gpreviewservice} only allows reviews of last week to be crawled for each app.
Instead, we built the comment dataset using two alternative approaches.
For the 8,400 apps we selected from the Chinese markets, we resort to a third-party app monitoring platform named Kuchuan\cite{kuchuan}, which has maintained the app metadata including comments from all the Chinese markets we studied.
For the 1,600 apps from Google Play, 
we developed an automated tool to continuously fetch the user comments \textit{everyday} within the span of 3 months.
Table \ref{table:comment} shows the distribution of collected comments. 
In total, we have collected over 3.2 million user comments from 5,156 apps\footnote{Note that, for the selected 10K app candidates, over 4,000 of them have no user comments or very few user comments, which were discard by us.}, including 192,982 comments from 2,027 malware, 1,960,728 comments from 1,416 (including 1,163 Chinese apps and 253 Google Play apps) grayware and 1,075,751 comments from 1,713 (including 1,157 Chinese apps and 556 Google Play apps) benign apps.
This dataset will be used in the large-scale measurement study (see \S~\ref{sec:measurement}).

\begin{table}[t]
\caption{Overview of our comment dataset.}
\label{table:comment}
\centering
\resizebox{\linewidth}{!}{
\begin{tabular}{l|cc|cc|cc}
\toprule
\multirow{2}*{Market} & \multicolumn{2}{c}{Malware} & \multicolumn{2}{c}{Grayware} & \multicolumn{2}{c}{Benign Apps} \\
\cline{2-7}
& \# apps &  \# comments & \# apps &  \# comments & \# apps &  \# comments \\
\midrule
360 Market~\cite{360market} & 625 & 33,432 & 399 & 205,383 & 457 & 161,286\\
Huawei~\cite{huawei} & 144 & 11,193 & 388 & 212,452 & 296 & 84,221 \\
Lenovo~\cite{lenovo} & 184 & 4,545 & 252 & 34,897 & 225 & 23,976 \\
Meizu~\cite{meizu} & 232 & 6,766 & 256 & 181,212 & 201 & 139,662 \\
Oppo~\cite{oppo} & 134 & 16,765 & 163 & 503,574 & 94 & 76,295 \\
Vivo~\cite{vivo} & 196 & 18,894 & 266 & 211,453 & 295 & 85,996 \\
Xiaomi~\cite{xiaomi} & 297 & 32,343 & 111 & 177,852 & 64 & 60,571 \\
Tencent Myapp~\cite{tencentmyapp} & 1117 & 69,044 & 477 & 250,649 & 481 & 131,949 \\
Google Play~\cite{googleplay} & NA & NA & 253 & 183,256 & 556 & 311,795 \\
\hline
Total & 2,027 & 192,982 & 1,416 & 1,960,728 & 1,713 & 1,075,751 \\
\bottomrule
\end{tabular}
}
\end{table}

\begin{figure*}[t]
\centering
  \includegraphics[width=0.8\textwidth]{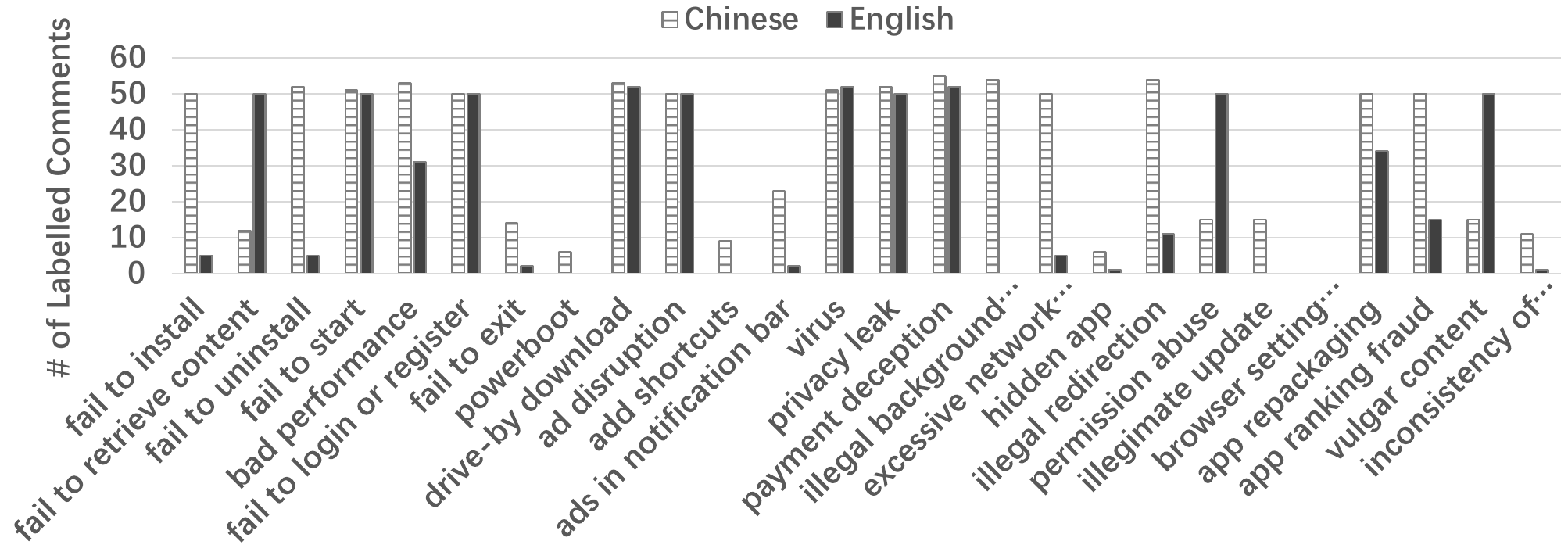}
\caption{Distribution of labelled benchmarks.}
\label{fig:benchmarkdata}
\end{figure*}

\begin{table*}[t]
\caption{Evaluation results on the benchmark datasets (best results are shown in bold).}
\label{table:benchmark}
\centering
\resizebox{\linewidth}{!}{
\begin{tabular}{c|c|c|c|c|c|c|c|c|c|c|c|c|c}
\toprule
\multirow{3}*{Category} & \multirow{3}*{Behavior} & \multicolumn{6}{c}{Benchmark (Chinese)} & \multicolumn{6}{c}{Benchmark (English)} \\
\cline{3-14}
& & \multicolumn{3}{c}{\tool} & \multicolumn{3}{c}{Similarity-Based Tool} & \multicolumn{3}{c}{\tool} & \multicolumn{3}{c}{Similarity-Based Tool}\\
\cline{3-14}
& & precision & recall & F1 & precision & recall & F1 & precision & recall & F1 & precision & recall & F1\\
\midrule
\multirow{8}*{\tabincell{l}{Functionality and \\ Performance}} & fail to install & \textbf{96\%}
 & \textbf{94\%} & \textbf{95\%} & 87\% & 90\% & 88\% & \textbf{83\%} & \textbf{100\%} & \textbf{91\%} & 71\% & 100\% & 83\%\\
~ & fail to retrieve content & \textbf{100\%} & \textbf{100\%} & \textbf{100\%} & 89\% & 67\% & 76\% & \textbf{96\%} & \textbf{98\%} & \textbf{97\%} & 85\% & 80\% & 82\%\\
~ & fail to uninstall & \textbf{100\%} & \textbf{96\%}  & \textbf{98\%} & 91\% & 92\% & 91\% & \textbf{100\%} & \textbf{100\%} & \textbf{100\%} & 63\% & 100\% & 77\%\\
~ & fail to start (e.g., crash) & \textbf{98\%} & \textbf{96\%}  & \textbf{97\%} & 88\% & 88\% & 88\% & \textbf{96\%} & \textbf{96\%} & \textbf{96\%} & 85\% & 82\% & 84\%\\
~ & bad performance (e.g., no responding) & \textbf{88\%} & \textbf{94\%}  & \textbf{91\%} & 86\% & 91\% & 88\% & \textbf{91\%} & \textbf{97\%} & \textbf{94\%} & 80\% & 77\% & 79\%\\
~ & fail to login or register & \textbf{98\%} & \textbf{98\%}  & \textbf{98\%} & 87\% & 90\% & 88\% & \textbf{96\%} & \textbf{100\%} & \textbf{98\%} & 87\% & 90\% & 88\%\\
~ & fail to exit & \textbf{93\%} & \textbf{93\%}  & \textbf{93\%} & 81\% & 93\% & 87\% & \textbf{100\%} & \textbf{100\%} & \textbf{100\%} & 100\% & 100\% & 100\%\\
~ & powerboot & \textbf{83\%} & \textbf{83\%}  & \textbf{83\%} & 83\% & 83\% & 83\% & NA & NA & NA & NA & NA & NA\\
\midrule
\multirow{4}*{\tabincell{l}{Advertisement}} & drive-by download & \textbf{100\%} & \textbf{94\%}  & \textbf{97\%} & 75\% & 85\% & 80\% & \textbf{100\%} & \textbf{98\%} & \textbf{99\%} & 73\% & 73\% & 73\%\\
~ & ad disruption & \textbf{100\%} & \textbf{100\%}  & \textbf{100\%} & 73\% & 64\% & 68\% & \textbf{100\%} & \textbf{100\%} & \textbf{100\%} & 69\% & 70\% & 69\%\\
~ & add shortcuts in launching menu & \textbf{100\%} & \textbf{100\%}  & \textbf{100\%} & 100\% & 78\% & 88\% & NA & NA & NA & NA & NA & NA\\
~ & ads in notification bar & \textbf{96\%} & \textbf{96\%}  & \textbf{96\%} & 55\% & 96\% & 70\% & \textbf{100\%} & \textbf{100\%} & \textbf{100\%} & 50\% & 100\% & 67\%\\
\midrule
\multirow{10}*{\tabincell{l}{Security}} & virus & \textbf{100\%} & \textbf{98\%} & \textbf{99\%} & 100\% & 86\% & 93\% & \textbf{100\%} & \textbf{100\%} & \textbf{100\%} & 100\% & 88\% & 94\%\\
~ & privacy leak & \textbf{98\%} & \textbf{94\%} & \textbf{96\%} & 88\% & 85\% & 86\% & \textbf{96\%} & \textbf{96\%} & \textbf{96\%} & 85\% & 82\% & 84\%\\
~ & payment deception & \textbf{100\%} & \textbf{91\%} & \textbf{95\%} & 92\% & 87\% & 90\% & \textbf{98\%} & \textbf{96\%} & \textbf{97\%} & 91\% & 79\% & 85\%\\
~ & \tabincell{c}{illegal background behavior (e.g., sms)} & \textbf{91\%} & \textbf{91\%} & \textbf{91\%} & 73\% & 76\% & 75\% & NA & NA &NA & NA & NA & NA\\
~ & excessive network traffic & \textbf{98\%} & \textbf{98\%} & \textbf{98\%} & 90\% & 90\% & 90\% & \textbf{100\%} & \textbf{100\%} & \textbf{100\%} & 80\% & 80\% & 80\%\\
~ & hidden app & \textbf{100\%} & \textbf{100\%} & \textbf{100\%} & 100\% & 67\% & 80\% & \textbf{100\%} & \textbf{100\%} & \textbf{100\%} & 100\% & 100\% & 100\%\\
~ & illegal redirection & \textbf{88\%} & \textbf{85\%} & \textbf{87\%} & 88\% & 78\% & 82\% & \textbf{92\%} & \textbf{100\%} & \textbf{96\%} & 75\% & 82\% & 78\%\\
~ & \tabincell{c}{permission abuse} & \textbf{92\%} & \textbf{80\%} & \textbf{86\%} & 86\% & 80\% & 83\% & \textbf{100\%} & \textbf{96\%} & \textbf{98\%} & 89\% & 84\% & 87\%\\
~ & illegitimate update (e.g., update to other app) & \textbf{87\%} & \textbf{87\%} & \textbf{87\%} & 86\% & 80\% & 83\% & NA & NA & NA & NA & NA & NA\\
~ & browser setting alteration & NA & NA & NA & NA & NA & NA & NA & NA & NA & NA & NA & NA\\
\midrule
\multirow{2}*{\tabincell{c}{Illegitimate Behavior \\of Developers}} & app repackaging & \textbf{85\%} & \textbf{92\%} & \textbf{88\%} & 78\% & 86\% & 82\% & \textbf{97\%} & \textbf{100\%} & \textbf{99\%} & 65\% & 65\% & 65\%\\
~ & app ranking fraud & \textbf{96\%} & \textbf{98\%} & \textbf{97\%} & 83\% & 80\% & 82\% & \textbf{83\%} & \textbf{100\%} & \textbf{91\%} & 69\% & 73\% & 71\%\\
\midrule
\multirow{2}*{\tabincell{l}{Content}} & vulgar content (e.g., pornography, anti-society) & \textbf{100\%} & \textbf{87\%} & \textbf{93\%} & 85\% & 73\% & 79\% & \textbf{100\%} & \textbf{92\%} & \textbf{96\%} & 95\% & 84\% & 89\%\\
~ & inconsistency between functionality and description & \textbf{100\%} & \textbf{91\%} & \textbf{95\%} & 83\% & 45\% & 59\% & \textbf{100\%} & \textbf{100\%} & \textbf{100\%} & 33\% & 50\% & 40\%\\
\bottomrule
\end{tabular}
}
\end{table*}

\section{Evaluation of \tool}
\label{sec:experiment}

\subsection{Benchmark Datasets}
\label{sec:benchmark}

We curated two benchmark datasets (English and Chinese) to evaluate \tool. We first select the apps which are confirmed to have undesired behaviors in the training dataset (see \S~\ref{sec:trainingdataset}). For each app, we exclude the comments already used in training dataset. At last, two authors of this paper manually inspected and labelled these comments.
Within our affordable efforts, we aim to collect and label 50 comments for each undesired behavior, except for some behaviors with few related apps.
Figure~\ref{fig:benchmarkdata} shows the distribution of our benchmark (901 Chinese comments and 618 English comments). 
Note that we cannot find comments for the behavior ``browser setting alteration''.

\subsection{RQ1: Effectiveness of \tool}

\subsubsection{Overall Results} Table~\ref{table:benchmark} shows the evaluation results.
\textbf{\emph{It shows that \tool is very effective in identifying \textit{UBComments}.}}
The average precision and recall are 95\% and 93\% for the Chinese benchmark, and 97\% and 98\% for the English benchmark. 
In particular, \tool achieves 90+\% of precision and recall for 20 out of 26 types of \textit{UBComments}.

\subsubsection{False Positives/Negatives} 
We further manually analyze the mis-classified comments and obtain two observations.
First, \emph{the false negatives are colloquial expressions instead of phrases.} 
For example, the comment ``A window of card application pops up continuously'' is describing the behavior ``ad disruption''.
But the key phrase ``ad'' is not in it. 
Moreover, if we add a new semantic rule with the phrases ``window'' or ``pop up'', it may lead to other false positives.
Second, \emph{the false positives are generated owing to our insufficiently conservative rules}. 
For example, the comment ``The app is completely useless, btw I thought that this built-in app can not be uninstalled, but it succeeded.'' is irrelevant to undesired behaviors.
However, it is classified to the behavior ``fail to uninstall'' since it has the phrases ``can not'' and ``uninstall''. 
Analogously, if we upgrade our rules to be more conservative, it may lead to more false negatives. 
These are the inherent limitations of rule-based matching methods.
We will further discuss it in \S~\ref{sec:discussion}.

\subsubsection{Comparison with Text Similarity Approach}
We compare \sys with the text similarity approach,
which classifies a comment to a type of undesired behavior based on text similarity between the comment and the classified comments in the training dataset (see \S~\ref{sec:trainingdataset}). 
We regard the behavior with the highest similarity score as the classification result.

As shown in Table~\ref{table:benchmark},
\sys achieves significantly better results than the text similarity approach. 
The average precision and recall achieved by the text similarity approach are 85\% and 81\% (v.s. 95\% and 93\% achieved by \tool) for the Chinese comment dataset, and 77\% and 85\% (v.s. 97\% and 98\% achieved by \tool) for the English comment dataset, respectively.
In particular, \sys outperforms the text similarity approach on all behaviors.
Such results indicate that the order and distance constraints adopted by our semantic rules can greatly reduce the false positives/negatives. 
For example, the comment ``I can not install the app'' is similar to ``I installed but it can not help me back up files'' considering their text similarity, but they are describing different types of undesired behaviors.
\tool correctly distinguishes these two comments while the text similarity approach classifies both of them to the same type of undesired behavior.

\section{Large-scale Measurement Study}
\label{sec:measurement}

\begin{table}[t]
\caption{Distribution of \textit{UBComments} by Categories.}
\label{table:genaralcategory}
\centering
\resizebox{0.8\linewidth}{!}{
\begin{tabular}{c|c|c}
\toprule
Category & \#Comment (\%) & \#App (\%)\\ 
\bottomrule
\tabincell{c}{Functionality/Performance} & 57,541 (61\%)  & 1701 (70\%)\\
Advertisement & 22,885 (24\%) & 1023 (42\%)\\
Security  & 13,765 (15\%) & 1098 (45\%)\\
\tabincell{c}{Illegitimate Behavior} & 1,129 (1\%) & 173 (7\%) \\
Content & 536 (0.5\%) & 72 (3\%)\\
\bottomrule
Total & 94,028 & 2,440\\
\bottomrule
\end{tabular}
}
\end{table}

\subsection{RQ2: \textit{UBComments} in the Wild}

\subsubsection{Overall Results}
From the dataset we harvested (see \S~\ref{sec:dataset}), \textbf{\sys identifies 94,028 \textit{UBComments}, belonging to 2,440 apps (47\%)}. 
Each app has received 39 \textit{UBComments} from multiple users on average.
This indicates that \textit{UBComments} are prevalent in the mobile app ecosystem, and the users who are sensitive to those policy violations are willing to report them in the comments.
Table~\ref{table:genaralcategory} shows the distribution of \textit{UBComments} and apps across different categories of behaviors.
Over 61\% of the \textit{UBComments} and over 70\% of the corresponding apps were complained to have ``functionality and performance'' issues.
This shows that users are most sensitive to the issues that directly affect their uses of the apps.
\textbf{For the 26 behaviors we summarized, 25 of them could be perceived by users.}
The most popular behaviors of \textit{UBComment} are ``fail to start'', ``ad disruption'', and ``payment deception'', accounting for 79.4\% of the \textit{UBComments}.
Both ``fail to start'' and ``ad disruption'' are related to user experiences, while ``payment deception'' shows users' security concerns.

\textbf{Manual Verification of Undesired Behaviors.} 
To analyze whether the undesired behaviors described in user comments reflect the real behaviors of mobile apps, we make effort to perform a manual verification here. For each of the 25 identified perceived behaviors, we randomly select three apps (75 apps in total) and manually verify if indeed the apps violated the policies as described. 
Our manual verification follows a series of steps. We first install them on smartphones to see whether they have shown undesired behaviors as user complained (e.g., ad disruption and malicious behaviors, etc.). Then we rely on Testin~\cite{testin}, a service that provides app testing on thousands of real-world smartphones, to check the functionality and performance issues (e.g., fail to start and fail to install). Furthermore, we leverage static analysis tools (e.g., LibRadar~\cite{ma2016libradar} and FlowDroid~\cite{steven2014flowdroid}) to extract and inspect behavior-related app information (e.g., sensitive code, permissions and libraries). At last, for the apps in behavior ``app ranking fraud'', we compare their comments based on existing approaches proposed in \cite{hu2019want, spamming1} to find fake comments.
Overall, 72 apps (96\%) have been confirmed with the undesired behaviors as user commented. For the 3 unconfirmed cases (one in the vulgar content category, and two in the payment deception category), our dynamic analysis found that their services have stopped and our static analysis failed due to they have adopted heavy obfuscation and code protection using packing services.
Nevertheless, we show that \textit{most of the undesired behaviors can  be confirmed}.

\begin{table}[t]
\caption{Distribution of \textit{UBComments} by Rating Stars.}
\label{table:stardistribution}
\centering
\resizebox{0.99\linewidth}{!}{
\begin{tabular}{c|c|c|c|c|c}
\toprule
Dataset & 1-star & 2-star & 3-star & 4-star & 5-star \\ 
\midrule
Malware & 28.14\%  & 17.70\% & 9.09\% & 3.85\% & 2.16\%  \\\midrule
Chinese Grayware & 23.36\%  & 16.22\% & 9.23\% & 4.37\% & 0.64\%  \\\midrule
Chinese Benign Apps & 25.38\%  & 17.58\% & 10.07\% & 5.95\% & 0.96\%  \\\midrule
GPlay Grayware & \tabincell{c}{6.50\%\\(14.28\%)}  &  \tabincell{c}{0.05\%\\(0.07\%)} &  \tabincell{c}{0.03\%\\(0.04\%)} &  \tabincell{c}{0.02\%\\(0.02\%)} &  \tabincell{c}{0\%\\(0\%)} \\\midrule
GPlay Benign Apps & \tabincell{c}{1.96\%\\(8.64\%)}  & \tabincell{c}{0.33\%\\(0.92\%)} & \tabincell{c}{0.15\%\\(0.25\%)} & \tabincell{c}{0.07\%\\(0.10\%)} & \tabincell{c}{0.01\%\\(0.02\%)}  \\\midrule
Total & 19.45\% & 12.15\% & 6.69\% & 2.78\% & 0.71\%  \\
\bottomrule
\end{tabular}}
\end{table}

\subsubsection{Low-rating Comments vs. High-rating Comments (RQ2.1)}
We study the distribution of \textit{UBComments} across comments with different ratings (from 1-star to 5-star).

\textbf{Quantitative Analysis.} 
As shown in Table~\ref{table:stardistribution}, it is apparent that \textbf{low-rating comments (\ie 1-star and 2-star) are more likely to describe undesired behaviors}. \textit{UBComments} account for roughly 20\% and 12\% for the 1-star comments and the 2-star comments, and 2.78\% and 0.71\% for the 4-star and the 5-star comments, respectively.
Note that the proportion of \textit{UBComments} in Google Play is much lower than that of Chinese markets.
The major reason is that the crawled comments from Google Play contain a large amount of blank comments, \ie the comments with only a rating but no descriptions. We further eliminate such comments and report the result (see the percentage in brackets in Table~\ref{table:stardistribution}).

\begin{figure*}[t]
\centering
  \includegraphics[width=1\textwidth]{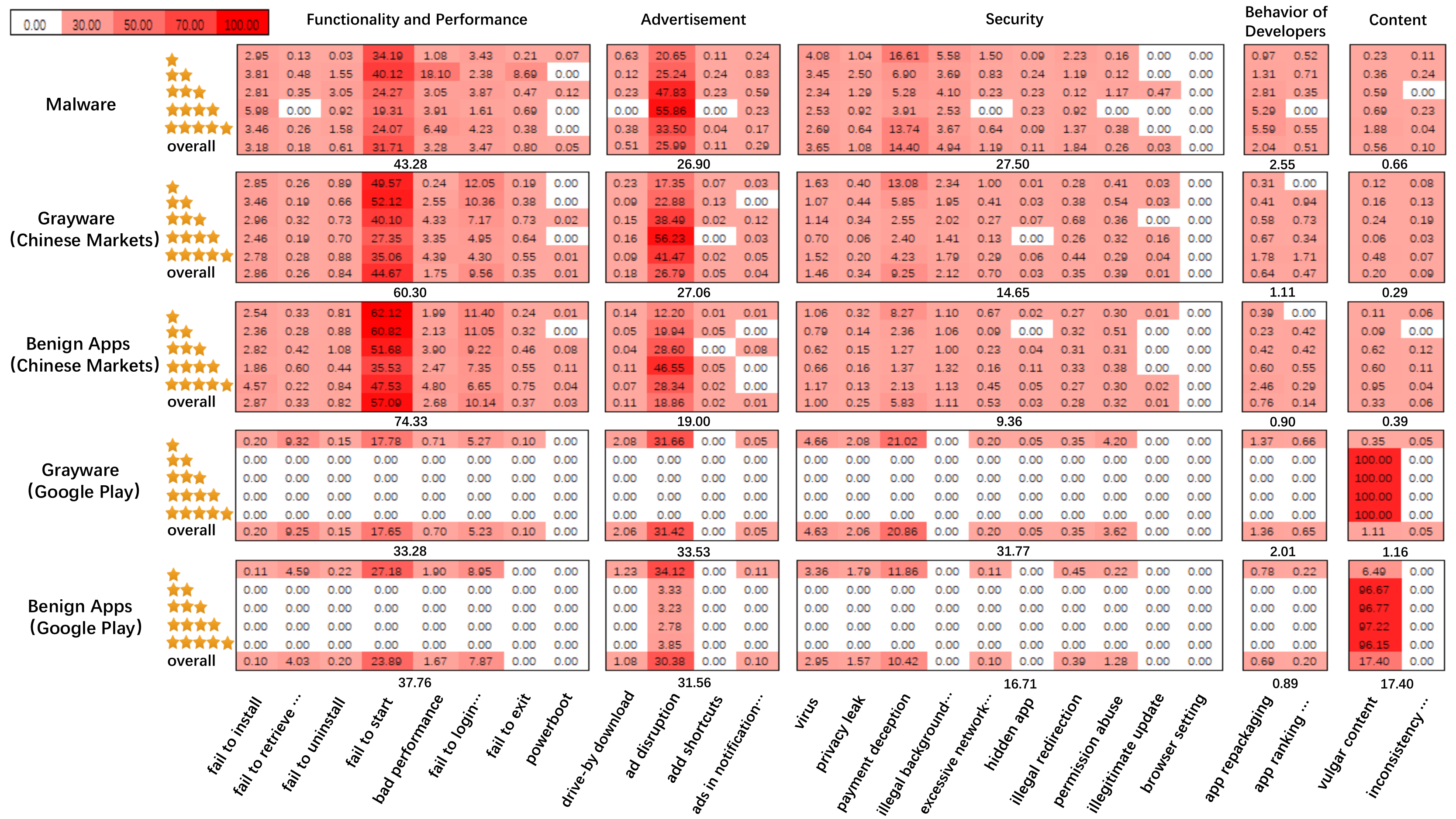}
\caption{Distribution of \textit{UBComments} across different ratings (y-axis) and undesired behaviors (x-axis) in different app categories.
Each row adds up to 100\%, with each cell representing the percentage of a specific undesired behavior in the \textit{UBComments} for a specific rating (\eg 5 stars) in an app category (\eg Malware). 
The depth of the color is also used to indicate the percentage in the cell, \ie, a deep color indicating a large percentage.
For each app category, a behavior category box represents the \textit{UBComments} of a specific behavior category (\eg Security), and the number below the box shows the percentage of the \textit{UBComments} in that behavior category.
}
\label{fig:measurement}
\end{figure*}

\textbf{Qualitative Analysis.}
As shown in Figure~\ref{fig:measurement}, the distributions of \textit{UBComments} in Chinese markets and Google Play show great diversity, and thus we discuss them separately.

For app comments in Chinese markets, the distribution of undesired behaviors does not show much diversity across \textit{UBComments} with different ratings.
Behaviors of the ``Functionality and Performance'' and ``Advertisement'' types are most prevalent across all the ratings,
with the ``Fail to start'' and ``Ad disruption'' types are quite noticeable.
Moreover, we find that security related behaviors are prevalent in both low-rating and high-rating comments of malware, but only prevalent in low-rating comments of grayware and benign apps.
It is quite surprising that users complain about the security issues (\eg payment deception) but give the app (malware) a high rating. 
Thus, we make efforts to manually examine all such ``contradictory'' comments (21,859 in total), and identify two major reasons. 
First, the default comment rating of most Chinese app markets is 5-star, thus a number of users may only complain the app in the comments but forget to assign a rating. 
Second, it is quite possible that some users misunderstand the meanings of 1-star and 5-star. 
For example, we find that several users assign totally opposite ratings in all their comments, \ie, 1-star with really good comments, but 5-star with negative comments, including the \textit{UBComments}.
It suggests the poor knowledge of the rating system for market users, and the new challenges in analyzing the comments of third-party app markets. 
Nevertheless, \tool can reveal how the users feel about their experiences, and even could improve the techniques of app risk assessments based on user comments~\cite{apprisk,autoreb}.

In Google Play, the distribution of \textit{UBComments} in low-rating comments and high-rating comments are quite different. Users generally give 1-star in their comments when they find undesired behaviors in the app, even if the behaviors do not belong to the ``security'' category. We only find a few comments that are related to the ``vulgar content'' type in other comments. 
This might be due to the high-quality market which pays more attention to policy regulations, and this more mature and regulated ecosystem enables users to better comprehend the ratings when providing comments.

\subsubsection{Malware vs. Grayware vs. Benign Apps (RQ2.2)}
For Chinese markets, over 42\% of malware samples have \textit{UBComments}, and they have occupied 7\% of the comments. 
As a contrast, over 57\% of benign app samples and 57\% of grayware samples have \textit{UBComments}, and the percentages of these comments are 4\% and 3\%, respectively.
For Google Play, over 32\% of benign apps and 38\% of grayware apps have \textit{UBComments}, and they account for 0.3\% and 1\% of the overall comments (0.6\% and 1.5\% after removing the empty ones from the overall comments), respectively.
In general, one would think that malicious apps have more UBComments than gray and benign apps, as their behaviors are more likely to inconsistent with users' expectation.
However, the results are different for what we expected, i.e., the percentage of \textit{UBComments} does not show much difference across malware, grayware and benign apps.
There are mainly two reasons. 
First, the policy-violation behaviors of two major types, ``Functionality and Performance'' and ``Advertisement'', are prevalent in both malicious and benign apps, \eg over 74\% of the UBComments in third-party benign apps are related to ``Functionality and Performance''. 
Second, some malware samples were removed in time by markets, and thus malicious apps have not received much complaints than expected.
Note that the security-related undesired behaviors show different distributions across malicious, gray, and benign apps (see Figure~\ref{fig:measurement}). 
As to Chinese markets, over 27\% of the \textit{UBComments} belong to the security category for malware, while the percentages for grayware and benign apps are 14\% and 9\%. 
As to Google Play, over 31\% of the \textit{UBComments} in grayware are security related (V.S. 16\% in benign apps).
Furthermore, we observe that \textbf{many user-perceived undesired behaviors (including security-related ones) were found in both malware and ``benign apps''. It suggests that some malicious behaviors are hard to detect by AV engines but user comments could provide insights for capturing them.}

\subsection{RQ3: Undesired Behaviors Across Markets}

We perform market-level analysis to investigate the differences across markets. 
On one hand, for the undesired behaviors declared in the policies of each market, we seek to measure how many such behaviors have been identified in our dataset. 
This result could be used to measure the effectiveness of market regulation, \ie how many of these undesired apps have bypassed the corresponding auditing process. 
On the other hand, for other undesired behaviors that were not declared in the policies of a market, we seek to explore whether we could find such behavior related comments in the corresponding markets. 

Table~\ref{table:undesiredbehavior} shows the results.
Roughly 34\% to 65\% of the apps (the numbers in bold) from each market have found comments for undesired behaviors described in each of their market policies. 
Over 65\% of the apps in Huawei Market have violated its market policies, while the percentage of such apps in Google Play is 34\%. 
From another point of view, roughly 5\% to 60\% of the apps (besides Google Play, as it covers all the behaviors we summarized in this paper) have been complained of having undesired behaviors that are not captured by the markets' policies. 
For example, over 60\% of the apps in the 360 Market have undesired behaviors that are not listed in its market policies. 
\textbf{This may open doors for malicious developers to exploit the insufficient vetting process.}

\section{Discussions}
\label{sec:discussion}

\subsection{Relation with Program Analysis}
\label{subsec:programanalysis}
A large number of papers were focused on using program analysis to detect the security~\cite{security3, gorla2014checking, fan2019graph}, privacy~\cite{wang2015using,liu2016identifying,pandita2013whyper, avdiienko2015mining,wang2017understanding}, ads/third-party library~\cite{ma2016libradar, dong2018frauddroid,wang2015wukong,liu2020maddroid}, and functionality issues~\cite{li2018cid, wei2018understanding, performance2, performance3,appsigning} of mobile apps.
In contrast, this paper focuses on a different perspective, i.e., \textit{how the users feel about their experiences}. 
Users’ expectations play a big role on how much the users can tolerate the apps’ behaviors.

First, although program analysis could be adopted to identify whether some sensitive behaviors exist in mobile apps, \textbf{it is non-trivial to verify whether the behaviors violate the policy. 
The borderline between policy-violation and tolerable misbehaviors is fuzzy and highly dependent on users’ subjective expectations}. 
For example, program analysis can easily identify ad libraries used in apps.
However, aggressive mobile ads cannot be simply conflated with the detection of ad libraries.
The detection of ad libraries, enabled by program analysis techniques, cannot take what users really feel about the ads into consideration. 
Second, \textbf{a number of the policy-violation behaviors, \eg payment deception and vulgar content, are difficult to be triggered and detected by program analysis techniques}.
However, they are indeed much easier revealed by user comments.

\textbf{Thus, \tool is complementary to program analysis, which can provide insight to identify the boundary between policy-violation behaviors and tolerable misbehaviors. }
Instead of identifying the policy-violation behaviors directly, \tool can serve as a \emph{whistle blower} that assigns policy-violation scores and identifies most informative comments for apps (\eg putting security  related comments at top).
Note that, not all the apps with UBComments should be removed by the app market. App vetting is aimed at promoting the overall quality of apps in the market. Thus, app markets would generally give developers warnings and buffer time to fix undesired behaviors in their apps (rather than removing them directly). With the help of \tool, it will be possible to pinpoint more urgent violations accurately, such as security-related ones, so that the markets could choose their reaction accordingly.

\subsection{Threats to Validity}
\emph{First, the taxonomy we summarized may be incomplete.} 
Although we have manually summarized 26 undesired behaviors, our taxonomy may still be incomplete since it was built based on current policies. 
However, our approach is generic and can be reused to support the detection of new types of undesired behaviors.
\emph{Second, our approach inherits the drawbacks of rule-based approaches.} Though our approach was proven to be quite effective during our evaluation, 
the semantic rules we summarized may not be complete and could introduce false positives/negatives as mentioned in Section~\ref{sec:benchmark}.
Nevertheless, market policies are rarely updated. Furthermore, our approach has strong expansibility of extracting new semantic rules for emerging app store policies. 
When policies evolve, new training can be performed to obtain new rules. Note that only the training process is semi-automated, as we need to manually label the classified comments. Our rules are extracted automatically from the labelled comments, which can be applied to identify UBComments automatically.
Third, \emph{we are not able to verify all the undesired behaviors for all the apps we identified}. We only sample 75 apps for manual verification, and found 96\% of them can be confirmed. 
We found \emph{most of the behaviors cannot be easily identified using automated tools, that is the reason why UBComments are prevalent even though these apps have already passed the market vetting process}. This motivates the research community to develop better tools for identifying such behaviors.
Nevertheless, as aforementioned (see Section~\ref{subsec:programanalysis}), instead of identifying the policy-violation behaviors directly, \tool could raise alarm based on the number of undesired comments and reported users.

\section{Related Work}

To the best of our knowledge, our paper is the first one that identifies undesired behaviors from user comments. Nevertheless, there are a number of studies focusing on app comments from different perspectives. 
We present and discuss briefly related works on (1) general app comment analysis, and (2) using NLP techniques in mobile app analysis.

\textbf{App Comment Analysis.}
Mobile app comments have been extensively studied from other perspectives, including mining user opinions~\cite{Khalid, Khalidarticle, Guzman, Galvis, fu2013, appreviewclassify8, rw7, rw9, rw10, Li-UbiComp17-compare, rw101, rw102,noei2019too}, app comment filtering~\cite{appreviewclassify6, fu2013, rw103}, and exploring other concerns~\cite{nguyen2019short, chen2013, cen2014, Palomba, chen2017}.
For example, Chen \textit{et al.}~\cite{appreviewclassify6} pioneered the prioritization of user comments with AR-Miner.
Chen \textit{et al.}~\cite{chen2013} conducted a study on the unreliable maturity content ratings of mobile apps, which will result in inappropriate risk exposure for the children and adolescents.
Nguyen \textit{et al.}~\cite{nguyen2019short} proposed to analyze the relationship between user comments and security-related changes in Android apps.
Kong \textit{et al.}~\cite{autoreb} presented a machine-learning technique to identify 4 pre-defined types of security-related comments.
Although app comments have been extensively studied from other perspectives, none of the above work correlates user comments to the undesired behaviors described in market policies and none of them can be easily adopted/extended to study this issue.

\textbf{NLP in Mobile App Analysis.}
Besides user comments, NLP techniques have been widely adopted to study app descriptions, privacy policies, and other meta text information related to mobile apps. 
Whyper~\cite{pandita2013whyper} and Autocog~\cite{qu2014autocog} adapt NLP techniques to characterize the inconsistencies between app descriptions and declared permissions. 
PPChecker~\cite{yu2016can} is a system for identifying the inconsistencies between privacy policy and the sensitive behaviors of apps. 
CHABADA~\cite{chabada} adapts NLP techniques to cluster apps using description topics, and then identifies the outliers of API usage within each cluster. 
Our work is the first to investigate the correlation between user comments and market policies.

\section{Conclusion}
We present the first large-scale study to investigate the correlation between user comments and market policies.
In particular, we propose \tool, a semantic-rule based approach that effectively identifies \textit{UBComments}. We apply \tool to a large scale user comment dataset and observe that \textit{UBComments} are prevalent in the ecosystem, even though app markets explicitly declared their policies and applied extensive vetting.
\tool offers a promising approach to detect policy violations, so as to help market maintainers identify these violations timely and further improve the app vetting process.

\section*{Acknowledgment}
This work was supported by the National Natural Science Foundation of China (grant numbers 62072046, 61702045 and 61772042), NSF (CNS-1755772), and Hong Kong RGC Projects (No. 152223/17E, 152239/18E, CityU C1008-16G).

\balance
\bibliographystyle{IEEEtran}
\bibliography{main}

\end{document}